\documentclass[
aps,
reprint
]{revtex4-1}

\usepackage{graphicx}

\usepackage[utf8]{inputenc}
\usepackage[T1]{fontenc}
\usepackage{mathptmx}
\usepackage{etoolbox}
\usepackage{lineno}
\usepackage{amsmath}
\usepackage{float}
\usepackage{upgreek}

\draft

\usepackage[hidelinks]{hyperref}
\hypersetup{
    colorlinks,
    linkcolor={blue},
    citecolor={blue},
    urlcolor={black}
}
\begin{document}

\title{Radiation reaction effects on particle dynamics in intense counterpropagating laser pulses}

\author{Caleb Redshaw}
\email[]{credshaw@stanford.edu}
\author{Matthew R. Edwards}
\email[]{mredwards@stanford.edu}
\affiliation{Department of Mechanical Engineering, Stanford University, Stanford, California 94305, USA
}

\date{\today}

\begin{abstract}
In high-intensity laser-plasma interactions, particles can lose a substantial fraction of their energy by emitting radiation. Using particle-in-cell simulations, we study the impact of radiation reaction on the dynamics of an underdense plasma target struck by counterpropagating circularly polarized laser pulses. By varying the relative wavelengths and intensities of the pulses, we find a range of parameters where radiation reaction can detrap electrons from the interference beat wave. The resulting charge separation field and the dominant direction of ion expulsion are thus reversed by radiative effects. Based on the electron dynamics during the interaction, we estimate the bounds on the parameter regime where the reversal occurs. The bounds take the form of three simple inequalities which depend only on the wavelength, normalized vector potential, and pulse duration ratios of the two lasers as well as the product of the pulse duration with a dimensionless radiation reaction parameter. 
Our estimates, which predict whether radiation reaction will change the final ion direction for a given set of laser parameters, broadly agree with the simulated results.
Finally, we outline an experimental procedure by which the reversal could be used to observe the transition to radiation-dominated dynamics.
\end{abstract}


\maketitle 

\section{Introduction}
\label{sec:intro}
Radiation reaction is the process by which charged particles lose momentum and energy to the photons they emit. In intense fields, the radiative losses can be on the order of the particle energy, fundamentally altering particle motion \cite{Blackburn2020}. The classical radiation reaction parameter for an electron moving in a laser field, $r_c = \alpha a_0 \chi_e \sim \epsilon a_0^3$, represents the electron's radiated energy throughout a laser period relative to its overall energy\cite{Blackburn2020,Kirk2016RadiativeBeams,Piazza2008ExactWave}. Here, $\alpha\approx1/137$ is the fine structure constant; $a_0\equiv eE_0/m_e\omega c$ is the normalized vector potential of a laser field with electric field amplitude $E_0$ and frequency $\omega$; and $\chi_e \sim \gamma E/E_S$ is the electron quantum parameter, equal to the electric field that the particle feels in its rest frame divided by the Schwinger critical field strength \cite{Schwinger1951} $E_S\approx1.3\times10^{18}$ V/m. $E$ is the local electric field at the electron's position and $\gamma=\sqrt{1+(\mathbf{p}/m_ec)^2}$ is the Lorentz factor for an electron with momentum $\mathbf{p}$. The non-dimensional coefficient $\epsilon=\frac{4\pi}{3}\frac{e^2}{m_ec^2\lambda}$, which is proportional to the ratio of the classical electron radius $r_e \equiv \frac{e^2}{m_ec^2}$ to the laser wavelength $\lambda=2\pi c/\omega$, indicates the importance of radiation for an electron in an electromagnetic wave and determines the required $a_0$ for a desired $r_c$ \cite{Gonoskov2022,Bulanov2011Lorentz-Abraham-DiracSolutions,Koga2005NonlinearRegime,Bulanov2017ChargedPatterns}. The constants $m_e$, $e$, and $c$ are the electron mass, the elementary charge, and the speed of light, respectively. When the laser fields are sufficiently intense that $r_c\gtrsim0.1$, radiation reaction must be taken into account; if $r_c\gtrsim1$, radiation reaction dominates over the Lorentz force, marking the radiation-dominated regime \cite{Blackburn2020,Gonoskov2022,Hadad2010EffectsAcceleration}. As $\chi_e$ approaches 1, quantum effects become important as well and the radiation emission must be treated as discrete and stochastic rather than continuous \cite{Blackburn2020,Blackburn2024AnalyticalLasers}. More comprehensive discussions of radiation reaction and radiation-dominated dynamics can be found in the reviews in Refs.~\citenum{Blackburn2020},\citenum{Gonoskov2022}, and \citenum{Fedotov2023AdvancesFields}. 

Numerical and analytical studies have discovered interesting consequences of radiation reaction\cite{Tamburini2010RadiationAcceleration,Hadad2010EffectsAcceleration}. Notably, it can lead to continuous longitudinal acceleration of a charged particle by an electromagnetic wave \cite{Gunn1971OnWaves,Kirk2016RadiativeBeams}. 
Further, when radiation reaction is strong relative to the ponderomotive force, particles may no longer be accelerated away from ponderomotive potential maxima \cite{Fedotov2014RadiationEffect}. Radiation reaction has been experimentally studied by examining electron energy spectra following interactions between ultra-relativistic electron beams and intense lasers \cite{Poder2018,Cole2018}. In these experiments, the electron energy losses agreed best with semi-classical\cite{Poder2018} or quantum\cite{Cole2018} descriptions of the radiation.
Other signatures have also been proposed. For instance, radiatively damped electron motion in conjunction with electron-positron pair production can cause upshifts in the transmitted laser frequency \cite{Griffith2022ParticleSignatures,Qu2021SignatureCascades}. The stochasticity of quantum radiation reaction leads to potentially observable "straggling" and "quenching", wherein particles reach classically forbidden regions of phase space \cite{Blackburn2014QuantumCollisions,Harvey2017QuantumPulses}.  

Counterpropagating waves, which are a well-suited geometry for studying strong-field quantum electrodynamic (QED) phenomena\cite{Bell2008,Kirk2009}, have been the focus of several previous studies \cite{Sheng2002,Sheng2004,Blaclard2023,Zhang2019,Lv2021,Aliani2023InfluenceElectron,Baumann2016,Esirkepov2017ParadoxicalFriction,Tiwary2021ParticleFrequencies}
. Without radiation reaction, relativistic ($a_0 \geq 1$) counterpropagating lasers can lead to chaotic particle motion and stochastic heating \cite{Sheng2002,Sheng2004}. For high-$a_0$ waves of equal amplitude and frequency, the particle dynamics can be described by a random walk model (for chaotic systems) or by analogy to a pendulum (for non-chaotic systems, e.g. perfectly counterpropagating circularly polarized waves) \cite{Blaclard2023}. It has also been shown that if one wave is smaller in amplitude and much greater in frequency, the dominant wave amplitude and the frequency ratio between the waves dictates the motion of the most energetic electrons\cite{Zhang2019}. 

The classical electron dynamics in counterpropagating circularly polarized waves were analyzed in detail from a single particle motion perspective in Ref.~\citenum{Lv2021} (see also Ref.~\citenum{Aliani2023InfluenceElectron}). An electron's trajectory was found to depend on its drift momentum, which itself depends on the details of the laser turn-on process. Importantly, a wave co-propagating with the electron has a greater influence on its final momentum, and therefore electrons tend to move with whichever laser they encounter first. However, this analysis treated the case where the second laser arrives only after the electron feels the full amplitude of the first laser, rather than the case where the second laser arrives during the ramp-up time of the first laser.

With radiation reaction, the particle dynamics in counterpropagating waves change.  
Radiative damping can cause particles to lose energy and bunch together, leading to a contraction in phase space \cite{Tamburini2011, Lehmann2012, Gong2016}. Particles are thus trapped around electric field nodes in standing waves, mitigating (but not necessarily eliminating) any chaotic motion \cite{Gonoskov2014,Lehmann2012,Kirk2016RadiativeBeams,Bulanov2017ChargedPatterns,Esirkepov2015AttractorsWaves}. If radiation reaction is strong, the particles may approach limit cycles around the standing wave electric field nodes or near the electric field anti-nodes, or follow chaotic trajectories around strange attractors in the vicinity of the electric field nodes \cite{Bulanov2017ChargedPatterns,Esirkepov2015AttractorsWaves}. In linearly polarized fields with intensity $I\gtrsim 6\times10^{25}$ W/cm$^{\textrm{2}}$, particles can instead be trapped around electric field anti-nodes in a process called anomalous radiative trapping \cite{Gonoskov2014,Baumann2016,Esirkepov2017ParadoxicalFriction,Bulanov2017ChargedPatterns}. 

The behavior of electrons---comprising an initial seed population and a population created via electron-positron pair production---in intense colliding laser pulses of different frequency was studied in Ref.~\citenum{Tiwary2021ParticleFrequencies}. For pulses of equal $a_0$, it was shown that the electrons have, on average, zero longitudinal momentum in the zero-momentum frame (ZMF) in which the pulses have equal frequency. With respect to the laboratory frame, the ZMF moves at 
\begin{equation}\label{eq:vZMF}
    v_{ZMF} \equiv \frac{R_\lambda-1}{R_\lambda+1}\text{,}
\end{equation} (normalized by the speed of light $c$), where $R_\lambda\equiv\lambda_2/\lambda_1$ is the wavelength ratio of the pulses.
$v_{ZMF}$ is also the phase velocity of the beat wave formed by the interference of the two pulses. For pulses of differing $a_0$, it was proposed that the appropriate frame is instead the zero-power-flow frame, in which the powers transmitted by both pulses are equal and thus there is no net transfer of energy in the longitudinal direction \cite{Tiwary2021ParticleFrequencies}:
\begin{equation}\label{eq:vZPF}
    v_{ZPF} \equiv \frac{R_\lambda-R_a}{R_\lambda+R_a}.
\end{equation}
Here, $R_a \equiv a_2/a_1$ is the pulse $a_0$ ratio. We will find in Section \ref{sec:resultsbeat} that the electron dynamics in our configuration depend directly on the motion of the interference beat wave, so $v_{ZMF}$ is the relevant velocity.

\begin{figure}
    \includegraphics[scale=1.00]{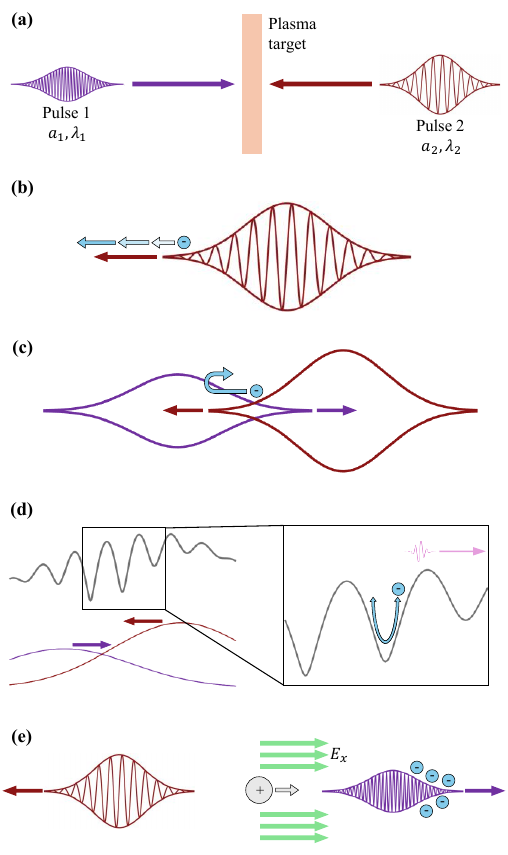}
    \centering
    \caption{\label{fig:schem} Schematic. (a) Circularly polarized laser pulses strike an initially stationary thin underdense plasma at normal incidence from both sides. The outer lines on each pulse represent the pulse envelope, while the interior sinusoidal curve represents one electric field component. (b) Longitudinal motion (blue arrows) of an accelerating electron (blue circle) under the influence of pulse 2 only. (c) Longitudinal motion of the same electron as it encounters pulse 1. The electron's longitudinal momentum is reduced to zero and it is turned around. (d) Standing wave electric field amplitude (gray curve, upper) resulting from the interference of pulses 1 and 2 (purple and red curves, lower) in the zero-momentum frame where $\lambda_1=\lambda_2$. The panel on the right shows the motion of a single electron trapped in the standing wave, oscillating between the sides of the potential well. As time proceeds, the pulses increase in amplitude. The pink sinusoid and arrow represent a high-energy emitted photon, corresponding to the radiative losses. (e) Longitudinal proton motion (gray circle and arrow) after the pulses have passed through the plasma. The protons are pulled by the charge separation field ($E_x$, green arrows) produced by the early electron dynamics. Here we show a representative case where the electrons moved with the beat wave, i.e. in the same direction as the higher-frequency pulse. The true direction in a given configuration depends on $R_a$ and $R_\lambda$, as well as whether radiation reaction is significant.}
\end{figure}

In this work, we investigate radiation reaction in the interaction between a thin underdense plasma and counterpropagating circularly polarized lasers of different intensities and wavelengths, as depicted in Fig.~\ref{fig:schem}(a). 
The laser pulses strike the plasma simultaneously, in contrast to the sequential configurations which have been studied in-depth in other works (including schemes in which electrons are first accelerated by a driver laser and then collide with a scattering laser) \cite{Cole2018,Poder2018,Lv2021,Griffith2022ParticleSignatures,Mirzaie2024AllOpticalLaser,Vranic2014AllOptical}. 
Through particle-in-cell simulations and simple single particle analysis, we find a regime in which radiation reaction reverses the particle motion during and after the interaction. 
We detail our physical configuration and one-dimensional (1D) computational parameters in Section \ref{sec:methods}. In Section \ref{sec:results}, we present the simulation results, including the electron dynamics within the laser pulses (Sections \ref{sec:resultsaccel}-\ref{sec:resultsbeat}) and the effect of radiation reaction on those dynamics (Section \ref{sec:resultsRR}). We provide a brief discussion of the 
resulting charge separation and proton motion in Section \ref{sec:resultsion}. Section \ref{sec:resultspond} discusses possible ponderomotive effects after the laser interaction. We then categorize the particle motion into five regimes based on the dynamics with and without radiation reaction, described in Section \ref{sec:resultsclasses}. In Section \ref{sec:2D}, we show that our results hold in two-dimensional (2D) simulations. Finally, in Section \ref{sec:discussion}, we suggest using the proton dynamics as a means to observe radiation reaction experimentally and summarize our findings.

\section{Methods}
\label{sec:methods}
To study the effects of radiation reaction, we have conducted particle-in-cell (PIC) simulations using \textsc{epoch} \cite{Arber2015ContemporaryModelling}. Select simulations were validated against \textsc{osiris} \cite{Fonseca2002OSIRIS,Hemker2000Particle-In-CellDimensions,Grismayer2017SeededPulses}, with no significant differences observed; the figures shown throughout this paper correspond to the \textsc{epoch} simulations.
The simulation geometry is shown in Fig.~\ref{fig:schem}(a). Short circularly polarized laser pulses are injected from both boundaries of a 1D domain. The pulse from the left (pulse 1 hereafter, denoted by subscript 1) has wavelength $\lambda_1=0.4$ \textmu m and normalized vector potential $a_1=100$. The pulse from the right (pulse 2 hereafter, denoted by subscript 2) has wavelength $\lambda_2=R_\lambda\times\lambda_1$ and normalized vector potential $a_2=R_a \times a_1$. Both pulses are Gaussian in time, with an intensity full width at half maximum (FWHM) of 10.6 fs except where specified otherwise. The pulses rotate with opposite senses as viewed along their propagation directions: pulse 1 is left-hand circularly polarized when looking in the $+\hat{\mathbf{x}}$ direction, while pulse 2 is right-hand circularly polarized when looking in the $-\hat{\mathbf{x}}$ direction. The domain spans 
a length of 66 \textmu m. A uniform, neutral electron-proton plasma with temperature $10$ eV and density $n_e\approx7\times10^{18}$ cm$^{-3}$ is initialized throughout the central 10 \textmu m. 
The simulations evolve for $\sim 60$ fs after the peaks of both pulse envelopes cross the center of the domain, allowing the particle dynamics to develop even after the two pulses pass each other. All simulations have a resolution of 40 cells per $\lambda_1$, and the initial plasma contains 40 particles per cell per species; resolutions up to 320 cells per $\lambda_1$ have been tested without meaningfully affecting our results.

Radiation reaction was included via \textsc{epoch}'s QED extension, which uses a Monte Carlo approach to model stochastic photon emission by nonlinear inverse Compton scattering \cite{Ridgers2014ModellingInteractions,Duclous2011MonteInteractions}. \textsc{epoch} simulates emission, and therefore radiation reaction, only for electrons and positrons. It is also capable of modeling nonlinear Breit-Wheeler electron-positron pair production\cite{Breit1934CollisionQuanta,Reiss1962AbsorptionLight}. However,
under the conditions of interest, the particle energies remain sufficiently low that $\chi_e\ll1$ and pair production is not significant. Additional simulations using \textsc{epoch}'s full QED module confirmed that pair production was negligible. 
We therefore only simulated the photon emission and resulting momentum loss. In order to directly see the impact of radiation reaction, we have performed each simulation both with and without the emission model enabled.

We have simulated conditions ranging from $1 \leq R_\lambda \leq 7$ and $1\leq R_a \leq 40$. Unless otherwise noted, the figures shown throughout this work correspond to simulations with $R_\lambda=3$ and $R_a=5$. The relevant computational parameters used for each figure are also provided in Appendix \ref{app:param}.
We have performed a limited set of 2D simulations with similar parameters, which we discuss in Section \ref{sec:2D}.

\section{Results}
\label{sec:results}

\begin{figure}
    \includegraphics[scale=1.05]{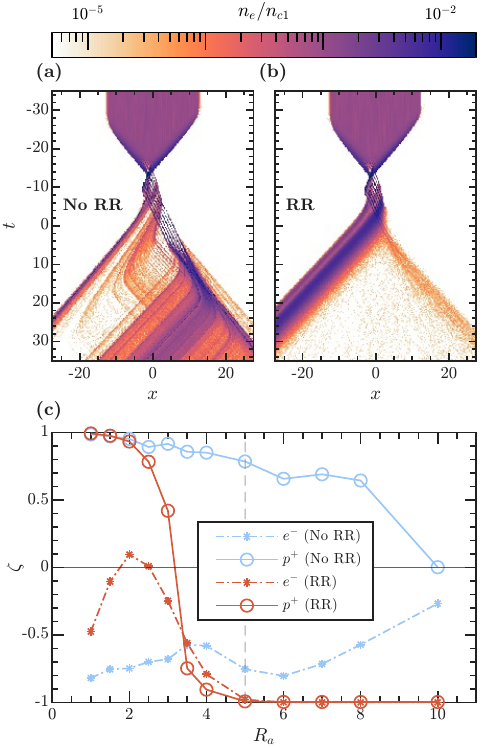}
    \centering
    \caption{\label{fig:netx} 1D simulation results. (a) Distribution of electron number density over time for simulations without radiation reaction. Pulse 1 ($\lambda_1=0.4$ \textmu m, $a_1=100$) is incident from the left, while pulse 2 ($R_\lambda=3$, $R_a=5$) is incident from the right. (b) Same conditions as (a), but with radiation reaction enabled in the simulation. The density evolution shows a clear change in favored direction compared to (a). (c) Results of a scan over $R_a$, showing $\zeta$ (measured at $t=45$, $\sim60$ fs after the pulse peaks pass each other) for electrons ($e^-$) and protons ($p^+$). The vertical dashed line indicates $R_a=5$, corresponding to the conditions of panels (a) and (b). There is evidently a range of $R_a$ for which $\zeta_{p^+}$ changes signs when radiation reaction is considered.}
\end{figure}

For certain parameters, we find that radiation reaction causes a reversal in the direction of particle motion. Example simulation results are shown in Fig.~\ref{fig:netx}(a) and (b). In Fig.~\ref{fig:netx} and hereafter, we normalize $x$ by $\lambda_1$ and $t$ by $T_1=\lambda_1/c$. The number densities shown are normalized by $n_{c1}=m_e\omega_1^2/4\pi e^2\approx7\times10^{21}$ cm$^{-3}$. Position $x=0$ corresponds to the center of the domain, and time $t=0$ corresponds to the instant when the pulse peaks coincide. When the pulses strike the plasma, the electrons initially feel only one pulse and are pushed towards the center (see Fig.~\ref{fig:schem}(b)). Once the pulses overlap, the electrons are decelerated (Fig.~\ref{fig:schem}(c)) and subsequently move with the interference beat wave (Fig.~\ref{fig:schem}(d)). As the system develops, the electrons may be detrapped from the beat wave. For the simulations shown in Fig.~\ref{fig:netx}, this occurs only if radiation reaction is included. The electrons then primarily move in one preferred direction (here, the $+\hat{\mathbf{x}}$ direction if they continue to move with the beat wave, or the $-\hat{\mathbf{x}}$ direction if they are detrapped). The electron motion produces a charge separation field that pulls the protons behind the electrons, as in Fig.~\ref{fig:schem}(e). 

To quantify the directionality, we define the quantity $\zeta$, measured after the interaction ends: 
\begin{equation}
    \zeta\equiv1-2f_{\text{left}}.
    \label{eq:zetadef}
\end{equation}
Here $f_{\text{left}}$ is the fraction of particles with momentum in the $-\hat{\mathbf{x}}$ direction. $\zeta=1$ means all particles are moving in the $+\hat{\mathbf{x}}$ direction, while $\zeta=-1$ means all are moving in the $-\hat{\mathbf{x}}$ direction. After the lasers pass, the proton dynamics are dominated by a slow expansion, which is directionally biased as a result of the early charge separation fields. The sign of $\zeta$, if measured a sufficient time after the interaction, reflects the long-term tendency of the system's motion. 
As we will see below, $\zeta_{p^+}$, i.e. $\zeta$ as measured considering protons only, is also a useful indicator of radiation reaction effects.
To measure $\zeta$, we integrate the particle $\hat{\mathbf{x}}$-momentum distribution functions output by \textsc{epoch} at the final timestep of each simulation. In these simulations, no protons leave the domain early, so the distribution functions account for all protons present.

Figure~\ref{fig:netx}(c) shows how $\zeta$ for both species varies over a range of $R_a$, keeping $R_\lambda=3$ fixed.
The protons show a large change in $\zeta$ when radiation reaction is included. However, the slow-moving protons do not directly experience radiation reaction under our conditions. Hence, it is the electron dynamics that drive the change; the slower proton response follows the initial electron response. We next examine the early electron motion, which is divided into three distinct stages: acceleration, particle stopping, and motion in the beat wave of the overlapping pulses.

\subsection{Initial acceleration}
\label{sec:resultsaccel}
\begin{figure*}
     \includegraphics[scale=1.05]{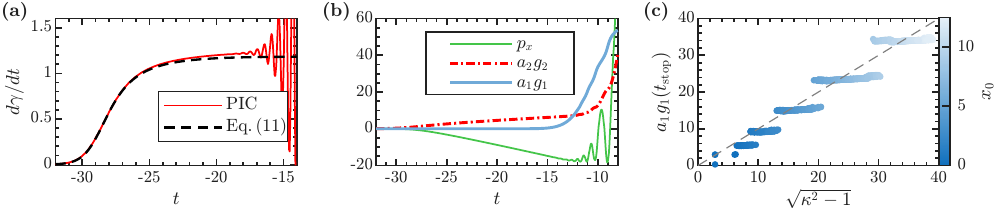}
     \centering
     \caption{\label{fig:SPM} (a) Average electron $d\gamma/dt$ vs time in the initial acceleration stage for $\sim 100$ macroparticles with initial positions $10.9\lesssim x_0 \lesssim 12.5$ from the simulation with $R_\lambda=3$, $R_a=5$, and radiation reaction disabled. Eq.~(\ref{eq:dgdt}) agrees with the PIC results until $t \gtrsim -18$, when the electrons begin to feel pulse 1. (b) Average local pulse normalized vector potentials $a_1g_1$ and $a_2g_2$ and normalized momentum $p_x$ over time for the same electrons as in (a). Once the electrons feel pulse 1, and especially during the stopping process ($-13\lesssim t \lesssim -10$, around the time when $a_1g_1$ becomes dominant), the amplitude of pulse 1 rises rapidly while that of pulse 2 rises by only about a factor of 2. (c) Numerically estimated $\sqrt{\kappa^2-1}$ for randomly sampled electrons across the right half of the plasma, compared to the PIC results for $a_1g_1$ at the time $t_{\text{stop}}$ when each electron's $p_x$ reaches 0. Color indicates the electron's initial position, with lighter color corresponding to a position further from the center. The dashed line represents a slope of 1, following Eq.~(\ref{eq:a1g1}). The electrons generally follow the predicted trend.}
\end{figure*}

We begin by considering a single electron starting at rest at the right edge of the plasma. Initially, it is influenced only by pulse 2 and accelerates in the $-\hat{\mathbf{x}}$ direction as illustrated in Fig.~\ref{fig:schem}(b). Because of the low $n_e$, any charge separation fields developed are negligible compared to the applied laser fields. The non-dimensional energy equation for the electron can be written as\cite{Gibbon} 
\begin{equation}
\frac{d\gamma}{dt}=\mathbf{v_\perp}\cdot\frac{\partial \mathbf{a_\perp}}{\partial t} = \frac{\mathbf{p_\perp}}{\gamma}\cdot\frac{\partial \mathbf{a_\perp}}{\partial t},
 \label{eq:energy}
\end{equation}
where $\mathbf{a}_\perp$ represents the transverse ($\hat{\mathbf{y}},\hat{\mathbf{z}}$) components of the local vector potential $\mathbf{a}$, and $\mathbf{v}_\perp$ and $\mathbf{p}_\perp$ are the transverse components of the particle's instantaneous velocity $\mathbf{v}$ and momentum $\mathbf{p}$ respectively. Bold-faced variables refer to vector quantities; unbolded variables of the same name (e.g. $a_{\perp}$) refer to their magnitudes. Here, we normalize $\mathbf{v}$ by $c$; $\mathbf{p}$ by $m_ec$; and $\mathbf{a}$ by $m_ec^2/e$. It can also be shown\cite{Gibbon} that
\begin{gather}
       \mathbf{p_\perp}=\mathbf{a_\perp}+\mathbf{p_{\perp,0}},\label{eq:pperp}\\
       p_x=-\gamma+\kappa,\label{eq:ppar1}
\end{gather}
and
\begin{equation}
   p_x=-\frac{p_\perp^2-\kappa^2+1}{2\kappa}\label{eq:ppar2},
\end{equation}
where $p_x=\mathbf{p}\cdot\hat{\mathbf{x}}$ denotes the $\hat{\mathbf{x}}$-component of $\mathbf{p}$, and $\kappa$ and $\mathbf{p_{\perp,0}}$ are constants that depend on the initial conditions. For an electron starting at rest, $\kappa=1$ and $\mathbf{p_{\perp,0}}=\mathbf{0}$. The negative sign in Eqs.~(\ref{eq:ppar1}) and (\ref{eq:ppar2}) appears because the wave is traveling in the $-\hat{\mathbf{x}}$ direction.
Substituting from Eq.~(\ref{eq:pperp}), Eq.~(\ref{eq:energy}) becomes
\begin{equation}
\frac{d\gamma}{dt}=\frac{\mathbf{a_\perp}}{\gamma}\cdot\frac{\partial \mathbf{a_\perp}}{\partial t}=\frac{1}{2\gamma}\frac{\partial a_\perp^2}{\partial t}. \label{eq:energy2}
\end{equation}  
For a circularly-polarized pulse with peak normalized vector potential $a_2$ and arbitrary envelope $g_2(x,t)$, $a_\perp=a_2g_2$ and the energy evolution follows 
\begin{equation}
    \frac{d\gamma}{dt} = \frac{a_2^2}{2\gamma}\frac{\partial}{\partial t}\left(g_2^2\right). \label{eq:dgdt_general}
\end{equation}
To proceed, we assume a Gaussian envelope with characteristic width $w$ $=\textrm{FWHM}/(\sqrt{2\ln2})$: $g_2(x,t)=e^{-\phi_2^2/w^2}$. Here, $\textrm{FWHM}$ refers to the FWHM of the associated intensity, $I_2(x,t)\sim a_2^2g_2^2(x,t)$. For an electron at position $x$ and time $t$, normalized as above, $\phi_2(x,t)=t+x-\phi_{d,2}$ is the local envelope phase, subject to the offset $\phi_{d,2}$ (which accounts for the initial separation between the particle and the pulse). Here, $w$ is normalized by $T_1$.
Equations (\ref{eq:pperp})-(\ref{eq:ppar2}) lead to $2\gamma-2=(a_2g_2)^2$, which can be inverted to find the envelope phase at which the electron will have a given $\gamma$ value:
\begin{equation}
    \phi_2 = -\frac{w}{\sqrt{2}}\sqrt{\ln\left(\frac{a_2^2}{2\gamma-2}\right)}, \label{eq:phi}
\end{equation}
where we retain only the negative root because the electron is on the rising edge of the envelope. Substituting the Gaussian envelope into Eq.~(\ref{eq:dgdt_general}), in conjunction with Eq.~(\ref{eq:phi}), gives a final ordinary differential equation for $\gamma$:
\begin{equation}
    \frac{d\gamma}{dt} = \frac{2\gamma-2}{\gamma}
    \frac{\sqrt{2}}{w}\sqrt{\ln\left(\frac{a_2^2}{2\gamma-2}\right)}. \label{eq:dgdt}
\end{equation}
Equation (\ref{eq:dgdt}) is straightforward to numerically evaluate, and describes the early-time electron energy evolution in our 1D PIC simulations. The agreement between Eq.~(\ref{eq:dgdt}) and our simulated results is shown in Fig.~\ref{fig:SPM}(a) for an average over $\sim100$ electron macroparticles that started near the right edge of the plasma. Deviation begins when the electrons start to feel pulse 1 at $t\approx-18$, leading to the next stage of motion.

\subsection{Particle stopping} \label{sec:resultsstop}
Once the electrons feel pulse 1---also Gaussian, with envelope $g_1(x,t)=e^{-\phi_1^2/w^2}$ and envelope phase $\phi_1(x,t)=t-x-\phi_{d,1}$---they begin to decelerate as shown schematically in Fig.~\ref{fig:schem}(c). Figure \ref{fig:SPM}(b) shows the instantaneous $p_x$ and the local normalized vector potentials of pulses 1 and 2 ($a_1g_1$ and $a_2g_2$ respectively), averaged over the same particles as in Fig.~\ref{fig:SPM}(a). It is clear that immediately after the electrons begin decelerating, $a_1g_1$ increases much more rapidly than $a_2g_2$ and to a much higher value. Because of this, we simplify the analysis by neglecting $a_2g_2$ during the stopping process. The second stage can then again be analyzed by following a single electron, now with initial momentum $p_0$ and moving in a single circularly-polarized wave with $a_\perp=a_1g_1$. From Eq.~(\ref{eq:ppar2}), we see that $p_x=0$, i.e.~the electron is longitudinally stopped, when $\kappa^2-1=p_\perp^2$. Since $p_\perp=a_\perp+p_{\perp,0}\approx a_1g_1$, this expression gives 
\begin{equation}
    a_1g_1(t_\text{stop}) \approx \sqrt{\kappa^2-1},  \label{eq:a1g1}
\end{equation}
where $a_1g_1(t_\text{stop})$ denotes the value of $a_1g_1$ at which the electron reaches $p_x=0$. 
The constant of motion $\kappa$ is set by the new initial conditions. From the acceleration stage, assuming that pulse 2 is sufficiently strong that the accelerated electrons become highly relativistic, $p_x\sim p_\perp^2\gg p_\perp$.
Thus we simplify by treating the initial momentum as purely in the $-\hat{\mathbf{x}}$ direction. Again using Eq.~(\ref{eq:ppar2}) and setting $p_x=-p_0$ and $p_\perp=0$ (with a sign change on the right-hand side, since our wave is now propagating in the $+\hat{\mathbf{x}}$ direction) we find
\begin{equation}
    \kappa \approx p_0+\sqrt{p_0^2+1}.  \label{eq:p0}
\end{equation}

To estimate $\kappa$, we numerically integrate Eq.~(\ref{eq:dgdt}) for starting positions $0<x_0<12.5$, matching the initial positions of electrons tracked within our PIC simulations. At each timestep, we update the positions using $dx/dt = p_x/\gamma = (1-\gamma)/\gamma$ and evaluate $a_1g_1$ and $a_2g_2$. We adopt the initial condition $\gamma_{ 0}=1+10^{-11}$ to avoid the singularity at $\gamma=1$. The value of $\gamma$ at the end of the first stage, estimated as the time when $a_1g_1=a_2g_2$, is then used to evaluate $p_0$ and, via Eq.~(\ref{eq:p0}), $\kappa$. Figure \ref{fig:SPM}(c) shows how Eq.~(\ref{eq:a1g1}), using $\kappa$ obtained in this way, compares to $a_1g_1(t_\text{stop})$ as extracted from the PIC results. The simulated results broadly follow the analytically-predicted trend. One interesting feature that arises is the stratification of the PIC results; rather than increasing steadily, the actual $a_1g_1(t_{\text{stop}})$ increases in intermittent jumps. This may be a result of the periodic interference between the two pulses, and suggests a limitation of the assumptions made in this analysis. However, this behavior is not essential for the underlying process of interest.

If $p_0$ is high enough, then Eqs.~(\ref{eq:a1g1}) and (\ref{eq:p0}) may predict a value for $a_1g_1(t_{\text{stop}})$ that is greater than $a_1$. Because $0\leq g_1 \leq 1$, a particle meeting this condition would not stop at all. From numerical tests, this occurs for very short pulses (FWHM $\lesssim7$ fs), small $a_1$ ($\lesssim 40$), large pulse $a_0$ ratios $R_a$ ($\gg 1$), or thicker plasmas (on the order of tens to hundreds of microns). Each of these constraints can be understood as follows. First, if the pulse duration is too low, a particle is accelerated more rapidly by the steep gradients, quickly reaching a momentum from which it cannot be stopped. Second, if $a_1$ is too low, then the assumption of highly relativistic motion no longer applies and our analysis of the stopping process is invalid. Third, if $R_a$ is too great, then pulse 2 remains the dominant pulse throughout, preventing the deceleration. Last, with a thick plasma, the particles towards the edges have longer to accelerate, again enabling them to reach values of $p_0$ that prevent stopping. Most of these restrictions do not pose a problem: the first is unrealistic to achieve in the near future, the second is undesirable for studies of radiation reaction, and the third necessitates impractically high laser power when paired with sufficiently high $a_1$. However, the need for thin plasmas is a practical concern for laboratory experiments (though it can be offset by using longer pulses---for pulses with FWHM $\approx30$ fs, $a_1=100$, and $R_a=5$, stopping is predicted for thicknesses close to 100 \textmu m). In the subsequent analysis, we assume that all particles do stop.

\subsection{Beat wave and detrapping} \label{sec:resultsbeat}
\begin{figure}
     \includegraphics[scale=1.05]{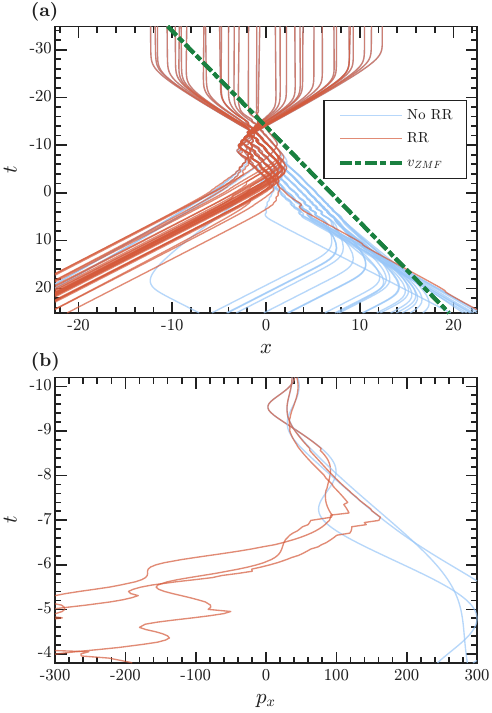}
     \caption{\label{fig:beatwave} (a) Individual electron trajectories without (blue) and with (orange) radiation reaction for the simulation with $R_\lambda=3$ and $R_a=5$. The dash-dotted green line corresponds to $v_{ZMF}$ (Eq.~(\ref{eq:vZMF})), which well-approximates the average velocity of the trapped electrons without radiation reaction. The oscillations seen in the trapped electron trajectories correspond to the oscillations within the potential wells as discussed in Section \ref{sec:resultsbeat}. The inclusion of radiation reaction does not substantially impact the trajectories until the trapping stage, at which point it enables the majority of particles to escape the beat wave. (b) Individual electron momenta over time from $-10\lesssim t \lesssim -4$ for three representative electron macroparticles, without (blue) and with (orange) radiation reaction. Radiation reaction leads to a decrease in $\hat{\mathbf{x}}$ momentum, nudging the electrons out of their trapped trajectories.}
\end{figure}

After coming to a stop, the electrons are trapped in the beat wave between the overlapping pulses and move, on average, at $v_{ZMF}$ in the $+\hat{\mathbf{x}}$ direction. Neglecting radiation for the moment, electrons will propagate with the beat wave until the pulses stop overlapping, unless the pulse dynamics can detrap them. The trapped beat wave motion can be seen in the blue electron trajectories plotted in Fig.~\ref{fig:beatwave}(a) from $-10\lesssim t \lesssim 6$. 

In the ZMF, the electrons are trapped in the stationary ponderomotive potential wells depicted in Fig.~\ref{fig:schem}(d). Assuming $a_2g_2>a_1g_1$, the right side of the well is higher than the left. A trapped particle, oscillating between the sides of the well, gains energy as the pulse amplitudes (and therefore the potential wells) grow in time. That particle may gain enough energy from the increasing field on the right side to subsequently escape when it traverses back to the left side, which is lower amplitude at any given time. Detrapping can therefore occur when the left-to-right variation of the peak amplitude exceeds some threshold of the temporal variation. The amplitude of the peaks is given by $a_1g_1+a_2g_2$. We thus set
\begin{equation} \label{eq:thresh}
C<\frac{\partial_x(a_1g_1+a_2g_2)}{\partial_t(a_1g_1+a_2g_2)},
\end{equation}
where the threshold $C$ is a numerical constant.

For the assumed Gaussian envelopes,
\begin{gather}
    \partial_xg_1 = \frac{2\phi_1}{w_1^2}g_1; \label{eq:derivs1} \\
    \partial_xg_2 = -\frac{2\phi_2}{w_2^2}g_2; \label{eq:derivs2} \\
    \partial_tg_1 = -\frac{2\phi_1}{w_1^2}g_1; \label{eq:derivs3} \\
    \partial_tg_2 = -\frac{2\phi_2}{w_2^2}g_2. \label{eq:derivs4}
\end{gather}

If the two pulses have equal $w$ in the laboratory frame, then in the ZMF they have width $w_1=w\sqrt{R_\lambda}$ and $w_2=w/\sqrt{R_\lambda}$.
The ratio in Eq.~(\ref{eq:thresh}) becomes
\begin{equation}
    C<\frac{-a_1g_1\phi_1/w_1^2+a_2g_2\phi_2/w_2^2}{a_1g_1\phi_1/w_1^2+a_2g_2\phi_2/w_2^2},
    \label{eq:thresh2a}
\end{equation}
or equivalently
\begin{equation}
    C<\frac{2a_2g_2\phi_2/w_2^2}{a_1g_1\phi_1/w_1^2+a_2g_2\phi_2/w_2^2}-1.
     \label{eq:thresh2}
\end{equation}
Moving the constant $-1$ into the threshold on the left-hand side and re-writing in terms of the pulse $a_0$ ratio $R_a$ and the pulse wavelength ratio $R_\lambda$ gives
\begin{equation}
    C<\frac{2R_aR_\lambda^2g_2\phi_2}{g_1\phi_1+R_aR_\lambda^2g_2\phi_2}.
     \label{eq:thresh3}
\end{equation}

\begin{figure*}
     \includegraphics[scale=1.05]{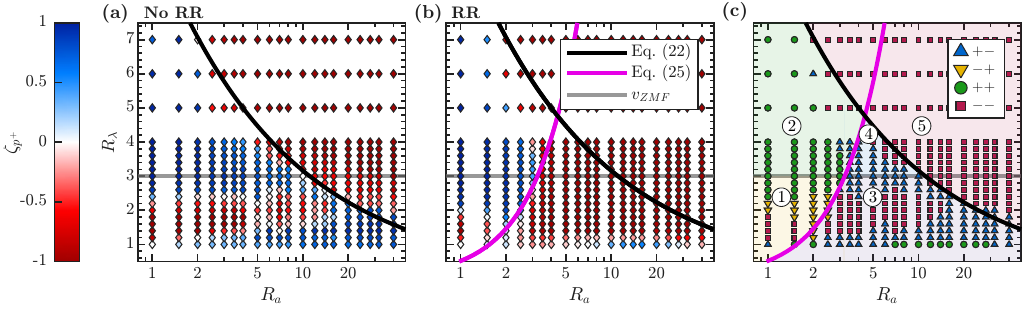}
     \centering
     \caption{\label{fig:bigscan} Simulation results for an $R_a$ and $R_\lambda$ parameter scan, showing $\zeta_{p^+}$ (a) without and (b) with radiation reaction. (c) Comparison of the sign of $\zeta_{p^+}$ across the scans. The two symbols ($+,-$) represent the sign of $\zeta_{p^+}$ at the corresponding points in panels (a) and (b), i.e. the blue upward triangle corresponds to $\zeta_{p^+}>0$ with no radiation reaction and $\zeta_{p^+}<0$ with radiation reaction. Regions corresponding to the interaction classes discussed in section \ref{sec:resultsclasses} are shaded and labeled with text bubbles. The black line in all panels represents Eq.~(\ref{eq:thresh4}) with $C=100$. 
     The horizontal gray line corresponds to $v_{ZMF}=1/2$. The pink line in panels (b) and (c) represents Eq.~(\ref{eq:rc1}).}
\end{figure*}

To obtain a global threshold which does not depend on the details of the particle positions,  we next assume that the electrons are concentrated around the center of the domain ($x \approx 0$) and the phase offsets are approximately equal ($\phi_{d,1} \approx \phi_{d,2}$). This allows us to cancel out the phases $\phi_1(x,t)$ and $\phi_2(x,t)$. We further assume that the envelopes $g_1(x,t)$ and $g_2(x,t)$ are of similar magnitude and can be canceled as well (as they are simply Gaussian envelopes from 0 to 1, and do not depend on $R_a$). Conceptually, this can be considered as accounting for the aggregate effect over the region where the pulses overlap, rather than for one specific trapped electron. This yields the approximate inequality $C\lesssim {2R_aR_\lambda^2}/({1+R_aR_\lambda^2})$, which can be rearranged to
\begin{equation}
    C\lesssim R_aR_\lambda^2.
     \label{eq:thresh4}
\end{equation}
Here, $C$ has been redefined to include all the constant factors. Note the higher-order dependence on $R_\lambda$, which implies that detrapping is increasingly difficult for lower wavelength ratios. When Eq.~(\ref{eq:thresh4}) is satisfied, most electrons are detrapped and swept along by wave 2. The resulting charge separation then sets up a longitudinal electric field which points along the $-\hat{\mathbf{x}}$ direction, driving the slow proton response. Fig.~\ref{fig:bigscan}(a) shows $\zeta_{p^+}$ for each simulation in a parameter scan over $1\leq R_\lambda \leq 7$ and $1 \leq R_a \leq 40$ (recall that $\zeta\equiv1-2f_{left}$ quantifies the direction of particle motion). The results of the scan agree with a threshold value $C=100$. 

Equation (\ref{eq:thresh4}) can be readily adapted to the more general case where the pulse durations are not equal in the laboratory frame. Consider pulses with laboratory-frame pulse widths $w_1^L$ and $w_2^L$, with pulse width ratio $R_w\equiv w_2^L/w_1^L$. In the ZMF, $w_1=w_1^L\sqrt{R_\lambda}$ and $w_2=w_2^L/\sqrt{R_\lambda}$. Using the fitted value for $C$, Eq.~(\ref{eq:thresh4}) is modified to
\begin{equation}
    C\approx100\lesssim\frac{R_aR_\lambda^2}{R_w^{2}}.
     \label{eq:thresh5}
\end{equation}
However, because the trapped velocity $v_{ZMF}$ depends only on $R_\lambda$ and is equal to 0 when $R_\lambda=1$, it is still desirable to have $R_\lambda>1$ in order for the electrons to preferentially move with the weaker pulse during the interaction.

\subsection{Detrapping with radiation reaction}
\label{sec:resultsRR}
We now consider radiation reaction. As seen in the orange trajectories of Fig.~\ref{fig:beatwave}(a), its inclusion produces little effect during either the initial acceleration or the particle stopping stage. This is not surprising, as the envelopes are far from their peaks during these stages and thus $r_c$ remains quite low. However, it changes the dynamics of the beat wave stage. As an electron propagates against the stronger pulse 2, photon emission reduces its energy and longitudinal momentum as represented in Fig. \ref{fig:schem}(d). If radiation reaction is strong enough, the electron is eventually freed from the beat wave.

The mechanism by which radiation reaction alters the single particle dynamics can be qualitatively understood in the following way. In the classical regime, radiation reaction on an electron can be approximated using the Landau-Lifshitz force \cite{Landau1975TheFields,Gong2016,Gonoskov2022,Bulanov2011Lorentz-Abraham-DiracSolutions}, which in the ultra-relativistic limit reduces to 
\begin{equation}
    \mathbf{F_{LL}}\approx-{\varepsilon}{\gamma^2}\mathbf{v}\left[(\mathbf{E}+\mathbf{v}\times\mathbf{B})^2-(\mathbf{v}\cdot\mathbf{E})^2\right].
    \label{eq:LL}
\end{equation}
Here, $\mathbf{F_{LL}}$ is the Landau-Lifshitz radiation reaction force, normalized by $m_ec\omega_1$; $\mathbf{E}$ and $\mathbf{B}$ are the local electric and magnetic fields, both normalized by $m_ec\omega_1/e$; and $\mathbf{v}$ is the electron's instantaneous velocity, normalized by $c$. It is clear that the dominant part of the radiation reaction acts opposite the particle's velocity \cite{Kirk2016RadiativeBeams}. 
The electrons are initially trapped within the beat wave (per Section \ref{sec:resultsbeat}), so their average longitudinal velocity is $+v_{ZMF}\hat{\mathbf{x}}$. This suggests that the radiation reaction force preferentially acts in the $-\hat{\mathbf{x}}$ direction when averaged over several oscillations. The net effect is that the electrons fail to keep up with the beat wave over time. In the quantum description, which better matches our simulation setup, the momentum is lost in the form of discrete high-energy photons, each providing a kick in the $-\hat{\mathbf{x}}$ direction. In either case, the electrons are able to escape the beat wave. In Fig. \ref{fig:beatwave}(b), we show the longitudinal momenta of a few representative trapped electrons from $-10\lesssim t \lesssim -4$. With radiation reaction, we see that each electron experiences a sudden drop in its momentum, attributable to photon emission, immediately before it breaks free of the beat wave. This observation supports the physical picture developed above: radiation indeed carries away momentum in the $+\hat{\mathbf{x}}$ direction, allowing electrons to leave the beat wave.

Since $r_c$ characterizes radiated energy over time and the radiation reaction mainly occurs during the pulse overlap time ($\sim w$), the total radiative losses during the interaction (relative to the particle energy) should depend on $r_c$ and $w$. Because $r_c\sim a_0^3$, radiation reaction is typically more significant for the higher-$a_0$ pulse 2. Therefore, we estimate $r_c$ using $a_2$ and $\lambda_2$. We take a simple approach by looking for particle detrapping along contours where the relative radiative losses should be approximately constant, i.e. contours of constant $r_cw/R_\lambda$. The factor of $R_\lambda$ is simply a result of our earlier choice of normalization: $w$ was normalized by $T_1$, but here we are concerned instead with its relationship to the period of pulse 2, $T_2=R_\lambda T_1$. Despite its simplicity, the radiative simulations largely agree with this estimate. The results of the same parameter scan as in \ref{sec:resultsbeat}, now including radiation reaction, can be seen in Fig.~\ref{fig:bigscan}(b). A threshold of 
\begin{equation}
    0.75\lesssim \frac{r_cw}{R_\lambda}
    \label{eq:rc1}
\end{equation} 
adequately describes the contour across which $\zeta_{p^+}$ changes from positive to negative. Because the protons follow the electron motion, we infer that if pulse 2 satisfies Eq. (\ref{eq:rc1}), it will detrap most electrons.
In terms of the pulse 1 parameters and the ratios $R_\lambda$ and $R_a$, this can be written as
\begin{equation}
    0.75\lesssim\frac{4\pi}{3}\frac{e^2a_1^3w}{m_ec^2\lambda_1}\frac{R_a^3}{R_\lambda^2}.
    \label{eq:rc2}
\end{equation}

One can gain insight into the subsequent behavior of the detrapped electrons by first considering the simpler case of a radiating particle in a single laser pulse. Radiation reaction induces a phase lag in such a particle's $\mathbf{v_\perp}$, which allows the electric field to do positive work on the particle throughout its trajectory; due to the laser's magnetic field, this work results in acceleration in the pulse propagation direction \cite{Gunn1971OnWaves,Kirk2016RadiativeBeams,Lehmann2011EnergyReaction}. For our particles, we again expect that radiation reaction is stronger for pulse 2 and the electrons will thus accelerate alongside that pulse. This expectation matches the orange trajectories of Fig.~\ref{fig:beatwave}: after pulse 2 frees the electrons from the beat wave, it subsequently pushes them in the $-\hat{\mathbf{x}}$ direction. In our simulations, we find that the acceleration generally follows the detrapping, i.e. if pulse 2 is sufficiently strong to detrap a particle, it is also sufficiently strong to accelerate it thereafter.

Note that, for the estimates in this section, we have used the classical radiation reaction parameter $r_c$ despite the fact that \textsc{epoch} uses a quantum model. For the key simulations where we observe a change in direction due to radiation reaction, $\chi_e \ll 1$ and our results are not significantly affected by the quantized nature of the emission.

\subsection{Charge separation and ion motion}
\label{sec:resultsion}
In our configuration, most electrons move in the same direction during the laser interaction. 
For the time being, we neglect the small population of electrons that exhibit different dynamics and assume that they essentially move uniformly. While the electrons are rapidly swept away from their initial positions, the protons respond on a slower timescale due to their greater mass. For $a_0\ll m_p/m_e\approx1836$ (where $m_p$ is the proton mass), the protons do not meaningfully move along the propagation axis as the laser pulses pass. The bulk motion of electrons alone thus creates an electrostatic charge separation field as depicted in Fig.~\ref{fig:schem}(e). 

The magnitude of the charge separation field depends on the displaced charge: $E_{ES}\sim n_eL$, where $E_{ES}$ is the electrostatic field and $L$ is a characteristic length scale for the displacement, assumed to be on the order of the plasma thickness. On the other hand, the peak amplitude of the combined laser fields scales as $a^*\sim a_1+a_2$, where $a^*$ is a normalized characteristic laser field. Comparing the two, and, for dimensional consistency, introducing some factors ignored above, we see that the laser field dominates over the electrostatic field by a factor $\sim a^*\times n_{c1}/n_e\times\lambda_1/L$. For all parameters explored in this work, this factor is much greater than unity. Therefore, during the laser interaction, the charge separation fields are indeed negligible and the electron motion is dominated by the laser fields (and radiation reaction, for $r_c\sim1$), which was previously assumed in Section \ref{sec:resultsaccel}.

The subsequent motion of the heavy protons can be approximated as purely due to the electrostatic fields. Continuing with the assumption of uniform motion, the expected behavior then largely follows that discussed in Ref.~\citenum{Bulanov2013StrongFoil}. In that work, which analyzes radiation pressure acceleration of a thin foil illuminated from a single side, the protons are first pulled after the electrons. After the electrons dephase from the laser pulse, they proceed to oscillate about the moving proton population, maintaining for some time an average velocity in the same direction as the initial charge separation\cite{Bulanov2013StrongFoil}. In our case, because the plasma density is so low, we do not see our electrons executing rapid electrostatic oscillations about the protons. Further, we do not expect our protons to be accelerated to particularly high energies, since the electrostatic fields driving them are weak. However, the primary qualitative feature---collective electrostatic fields, caused by electron motion and resulting in directional proton motion---remains applicable.

The electrostatic field development and resulting proton motion observed in our simulations are in fact more complex than the above description. Recall that the thresholds determined in Sections \ref{sec:resultsbeat} and \ref{sec:resultsRR} describe the overall electron behavior, rather than the behavior of any individual electron. Though the majority of electrons travel in the predicted direction, there are always some that move differently. As a result, the charge separation fields do not point in the predicted direction throughout the entire plasma. For example, if a small population of electrons remains trapped in the beat wave despite the pulses nominally being above the threshold in Eq.~\ref{eq:rc1}, that population will then act to pull a fraction of the protons in the $+\hat{\mathbf{x}}$ direction. Such deviations in the electron dynamics are the reason why the proton directionality as measured by $\zeta_{p^+}$ occupies a range of values between $\pm1$, rather than tending just towards the extremes. Intermediate values of $\zeta_{p^+}$ often occur for configurations near the reversal thresholds, which can be seen in Fig.~\ref{fig:bigscan}(a) and (b).

\subsection{Ponderomotive effects}
\label{sec:resultspond}
A third approximate criterion for an observable reversal can be determined by considering the ponderomotive effects on the electron motion at the end of the interaction. Here, we assume that the electron remains trapped during the interaction, i.e. none of Eq. (\ref{eq:thresh4})-(\ref{eq:rc1}) are satisfied.
During the pulse overlap time, the electrons trapped in the beat wave move forward an average distance that scales with $v_{ZMF}\times w$ while each pulse propagates forward by $w$. As the overlap ends, the electrons are subjected to a ponderomotive force which pushes them away from the envelope peaks. If $v_{ZMF}$ is close to 1, this is not an issue: most electrons will remain in front of the peak of pulse 1 and the ponderomotive force will act in the $+\hat{\mathbf{x}}$ direction, maintaining the charge separation that drives the preferential proton motion. As $v_{ZMF}$ decreases, more electrons fall behind the peak of pulse 1 and are instead immediately ejected backwards through the plasma. If enough electrons are ejected, the early charge separation field will pull protons in the $-\hat{\mathbf{x}}$ direction instead, potentially eliminating the desired reversal. In the limit $v_{ZMF}\to0$ ($R_\lambda\to1$), the electrons stay near the center of the domain and experience a ponderomotive push from the falling edges of both pulses as the overlap ends. For $R_w=1$, the gradients of the two pulse envelopes are nearly symmetric and the higher-$a_0$ pulse 2 provides the stronger ponderomotive force; the electrons thus move in the $+\hat{\mathbf{x}}$ direction. When $R_w\neq1$, the gradients are no longer symmetric, so the stronger ponderomotive force when $v_{ZMF}\approx0$ may come from either pulse depending on the local gradients.

To mitigate these ponderomotive effects and focus on the interplay between particle trapping and radiation reaction, we desire a high $v_{ZMF}$. However, from Fig.~\ref{fig:bigscan}(b), the distance between the two detrapping thresholds narrows as $R_\lambda$ (and therefore $v_{ZMF}$) increases, shrinking the range of $R_a$ for which a reversal occurs. $v_{ZMF}\gtrsim1/2$ appears to be a sufficient compromise, which works for all configurations tested herein. At this velocity, trapped electrons move in the same direction as pulse 1 but cover only half the longitudinal distance. An electron located at a position $\gtrsim w/2$ ahead of the peak of pulse 1 at the start of the beat wave stage remains ahead of the peak at the end of the overlap and avoids the ejection. This is not a precise limit, as the trapped electrons will in reality occupy a range of positions relative to the envelope peaks and some will always end up located such that they are pushed back through the plasma.  It is also overly restrictive in some cases, e.g. the $R_w=3/4$ simulations to be shown in Section \ref{sec:resultsclasses}, where the envelope gradients at moderate to low $v_{ZMF}$ tend to favor the $+\hat{\mathbf{x}}$ direction in which the electrons were already moving. However, it provides a useful guideline for estimating where the post-overlap ponderomotive force plays an important role.

In configurations where radiation reaction is present but not strong enough to detrap the particles, i.e.~$r_cw/R_\lambda<0.75$, the radiation reaction may still counteract the ponderomotive ejection. This can be thought of as a result of the electrons radiating away some of their backward and transverse momentum, similar to the findings in Refs.~\citenum{Fedotov2014RadiationEffect} and \citenum{Ji2014}. Weak radiation reaction can thus, for lower $v_{ZMF}$, actually enhance the early charge separation field in the $+\hat{\mathbf{x}}$ direction.

\subsection{Interaction regimes}
\label{sec:resultsclasses}
In Fig.~\ref{fig:bigscan}(c), we compare the sign of $\zeta_{p^+}$ between the non-radiative (Fig.~\ref{fig:bigscan}(a)) and radiative (Fig.~\ref{fig:bigscan}(b)) simulations. The comparison reveals five distinct regimes of interaction, bounded by the thresholds discussed in Sections \ref{sec:resultsbeat} and \ref{sec:resultsRR} as well as the approximate threshold $v_{ZMF} \gtrsim 1/2$ to avoid the ponderomotive effects discussed in Section \ref{sec:resultspond}. The interaction regimes are described below.
\begin{enumerate}
    \item \label{class1} \textit{No detrapping regardless of radiation reaction, and $v_{ZMF} \lesssim 1/2$}. During the overlap time, electrons co-propagate with the higher frequency pulse. Due to the low $v_{ZMF}$, they emerge behind the peak of $g_1$ and are immediately ejected back towards and through the plasma. Radiation reaction can mitigate this ejection. The resulting charge separation sets up a longitudinal electric field that directs the proton motion primarily in the $-\hat{\mathbf{x}}$ direction without radiation reaction, but in the $+\hat{\mathbf{x}}$ direction with it.
    \item\label{class2} \textit{No detrapping regardless of radiation reaction, and $v_{ZMF} \gtrsim 1/2$}. Electrons again co-propagate with the higher frequency pulse. Because of the high $v_{ZMF}$, they emerge ahead of the peak of $g_1$ and take much longer to fall behind and eject. The result is preferential proton motion in the $+\hat{\mathbf{x}}$ direction.
    \item\label{class3} \textit{Detrapping only with radiation reaction, and $v_{ZMF} \lesssim 1/2$}. Electrons emerge from the overlap with the stronger pulse only if radiation reaction is included. Without radiation reaction, they behave as in regime \ref{class1}. In either case, the proton motion is typically directed in the $-\hat{\mathbf{x}}$ direction. Note, however, that the reversal reappears at the bottom of this regime, as $R_\lambda$ approaches 1 and thus $v_{ZMF}$ approaches 0. This is also a result of the ponderomotive effects as discussed in Section \ref{sec:resultspond}.
    \item\label{class4} \textit{Detrapping only with radiation reaction, and $v_{ZMF} \gtrsim 1/2$}. Electrons behave as in regime \ref{class2} without radiation reaction, and as in regime \ref{class3} if it is included. Unlike regime \ref{class1}, the resulting charge separation field pulls the protons mostly in the $+\hat{\mathbf{x}}$ direction without radiation reaction and in the $-\hat{\mathbf{x}}$ direction with it.
    \item\label{class5} \textit{Detrapping regardless of radiation reaction}. Electrons in this case are always detrapped and so always move with the stronger pulse 2 in the $-\hat{\mathbf{x}}$ direction, as do the protons.
\end{enumerate} 

\begin{figure}
     \includegraphics[scale=1.05]{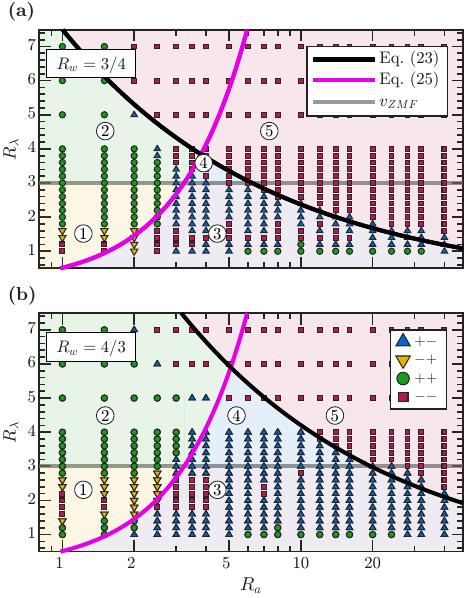}
     \centering
     \caption{\label{fig:wratio} Comparison of the sign of $\zeta_{p^+}$ across scans (a) with $R_w=3/4$ and (b) with $R_w=4/3$. The different symbols represent the sign of $\zeta_{p^+}$ with radiation reaction off and on, as in Fig.~\ref{fig:bigscan}(c). Regions corresponding to the interaction classes discussed in section \ref{sec:resultsclasses} are shaded and labeled with text bubbles. The black line represents Eq.~(\ref{eq:thresh5}), 
     while the pink line  represents Eq.~(\ref{eq:rc1}) using the minimum of $w_1^L$ and $w_2^L$. The horizontal gray line again corresponds to $v_{ZMF}=1/2$.}
\end{figure}

Figure \ref{fig:wratio} shows how the sign of $\zeta_{p^+}$ changes between simulations when the pulse durations are not equal, compared to the predicted bounds. In Fig.~\ref{fig:wratio}(a), $R_w=3/4$, while in Fig.~\ref{fig:wratio}(b), $R_w=4/3$. 
For the simulation scans shown in Fig.~\ref{fig:wratio}, we set the characteristic width of the shorter pulse---$w_1^L$ in Fig.~\ref{fig:wratio}(a), $w_2^L$ in Fig.~\ref{fig:wratio}(b)---to 6.75, corresponding to a pulse intensity FWHM of 10.6 fs. 
For Eq.~(\ref{eq:rc1}), we use the minimum $w$ between the two pulses, since the shorter pulse limits the overlap time. 
The prediction from Eq.~(\ref{eq:thresh5}) shifts towards higher $R_a$ when pulse 2 is longer than pulse 1, widening the gap between the two detrapping thresholds and therefore increasing the range of parameters for which we observe a reversal.

Exploratory simulations using pulses with $R_w=1$ and FWHMs of either $\sim14$ fs or $\sim18$ fs (not shown here) indicate that regimes \ref{class1} and \ref{class3} are suppressed for longer pulse durations. The increased overlap duration in longer pulses allows an early charge separation field to develop before the ejection occurs, giving the protons an initial pull in the $+\hat{\mathbf{x}}$ direction and so reducing the later impacts of the ponderomotive force. Because of this undesirable pulse duration dependence, regime \ref{class4} is instead the most likely candidate for experimental observation.

\subsection{2D PIC simulations}
\label{sec:2D}
To evaluate the utility of our results under more realistic conditions, we have also performed 2D simulations with $R_\lambda=3$ and $3\leq R_a \leq 7$, focusing on regime \ref{class4}. The 2D domain spans $-62.5\leq x\leq62.5$ (50 \textmu m total length) in the longitudinal direction, with the transversely uniform plasma located between $-12.5\leq x\leq12.5$ (again 10 \textmu m). Both laser pulses are transversely Gaussian and focused at the center of the domain to a beam waist varying from $4.25-42.5\lambda_1$ (corresponding to 1.7-17 \textmu m). At focus, $a_1=100$. Both beams are normally incident on the plasma target. The transverse boundaries are open, and are sufficiently wide ($y_{max/min}=\pm5$ times the size of the largest laser radius at the boundary) that their impact on the interaction is negligible. To reduce computational cost, the plasma is initialized with only 5 particles per cell; simulations with more macro-particles show minimal differences. All other parameters are the same as for the corresponding 1D simulations. 

\begin{figure}
     \includegraphics[scale=1.05]{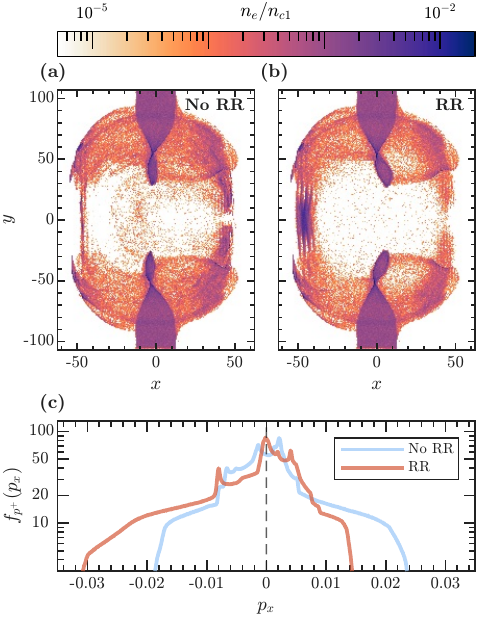}
     \centering
     \caption{\label{fig:2Dexample} Electron number density at $t=45$ in the 2D PIC simulations with $R_\lambda=3$, $R_a=5$, and a beam waist of 8.5 \textmu m, (a) without and (b) with radiation reaction. The same change in direction from the 1D simulations is present. (c) Proton momentum distribution functions $f_{p^+}(p_x)$ at $t=265$ in the longer 2D PIC simulations, again with $R_\lambda=3$ and $R_a=5$. The proton momentum, particularly in the tails of the distributions, is preferential towards the $-\hat{\mathbf{x}}$ direction when radiation reaction is included and towards the $+\hat{\mathbf{x}}$ direction when it is not.}
\end{figure}

We find that the change in direction is reproduced for large beam waists---4.25 \textmu m is sufficient for the reversal to appear under our conditions, though the signature is stronger with a waist of 8.5 \textmu m. For more tightly focused pulses, transverse ejection weakens the beat wave motion. Panels (a) and (b) of Fig.~\ref{fig:2Dexample} show the final ($t = 45$) electron density distributions in space for the simulations with $R_\lambda=3$, $R_a = 5$, and beam waist 8.5 \textmu m, without and with radiation reaction respectively. There is a clear difference in the electron behavior, with bunches propagating in the $-\hat{\mathbf{x}}$ direction for the radiative simulation which are absent otherwise. In these simulations, $\zeta_{p^+}\approx$ $0.2$ for (a) and $\zeta_{p^+}\approx-0.34$ for (b). This signature is weaker than the corresponding 1D signatures ($\zeta_{p^+}\approx 0.8$ and $\zeta_{p^+}\approx -1$ respectively). However, some of the reduction can be attributed to the many particles away from the center of the domain which are only weakly affected by the charge separation field and thus bring the overall $\zeta_{p^+}$ closer to 0. The effect persists for small deviations from normal incidence: in an otherwise equivalent 2D simulation with both pulses tilted by 5\textdegree\, relative to the target normal, $\zeta_{p^+} \approx 0.14$ without radiation reaction and $\zeta_{p^+} \approx -0.35$ with it.

Finally, to track the proton dynamics over a longer time, we have conducted 2D (normal incidence) simulations with $R_\lambda=3$ and $R_a=5$ until $t = 265$, $\sim350$ fs after the pulse peaks cross $x=0$. 
In Fig.~\ref{fig:2Dexample}(c), we show the resulting proton momentum distributions $f_{p^+}(p_x)$ at $t=265$, normalized so that they have unit integrals. The proton momentum is normalized by $m_pc$.
The early difference in $\zeta_{p^+}$ manifests as different directional biases in the later motion of the proton population left behind. The bias is particularly noticeable for the simulation with radiation reaction, for which the momentum distribution deviates more significantly from symmetry about $p_x=0$. Notably, the tails of the distributions extend to higher magnitudes in different directions: the left-moving particles attain greater momenta with radiation reaction and the right-moving particles attain greater momenta without it. The $\zeta_{p^+}$ signature itself diminishes from the early measurements, though it does not vanish. For the data in Fig.~\ref{fig:2Dexample}(c), $\zeta_{p^+}\approx$ $0.04$ and $\zeta_{p^+}\approx-0.13$ without and with radiation reaction respectively.

\section{Discussion} \label{sec:discussion}

The effect discussed herein suggests a straightforward experimental diagnostic by simply counting protons ejected after the laser interaction. 
Because $r_c$ depends strongly on $a_0$, the onset of radiation reaction could be observed experimentally by selecting a set of parameters slightly below the radiative threshold of Eq. (\ref{eq:rc1}) and then increasing $a_1$ until the reversal occurs (while holding the pulse $a_0$ ratio $R_a$ constant). For example, one could use two Ti:Sapphire lasers ($\lambda\approx 0.8$ \textmu m) at near-normal incidence to strike a 10 \textmu m-scale underdense plasma, itself formed by the expansion of a thin foil target ionized by the laser pre-pulses. To achieve the desired pulse wavelength ratio $R_\lambda=3$, pulse 1 would be frequency tripled to $\lambda_1\approx0.267$ \textmu m. For this wavelength, when $R_\lambda=3$ and $R_a=2.75$ and assuming an intensity FWHM of 10.6 fs as used in our simulations, particles are expected to move with the higher-$a_0$ pulse for $a_1\gtrsim90$ ($r_cw/R_\lambda\approx0.75)$ if radiation reaction is present. 

\begin{figure}
    \includegraphics[scale=1.05]{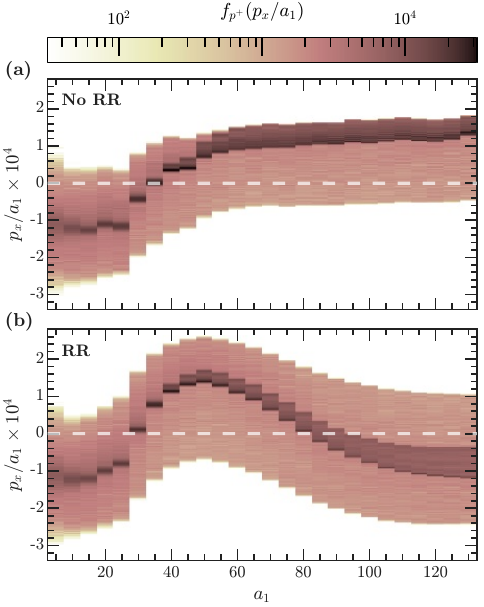}
    \centering
     \caption{\label{fig:exp_alt} Normalized proton momentum distribution functions $f_{p^+}(p_x/a_1)$ from the 1D PIC simulations of the experimental conditions detailed in Section \ref{sec:discussion}, (a) without and (b) with radiation reaction. Momenta are measured 60 fs after the pulse peaks pass each other. The white dashed lines indicate $p_x=0$. When $a_1$ (and therefore $r_c$) is greater, the proton momenta shift towards negative values if radiation reaction is included.}
\end{figure}

In Fig.~\ref{fig:exp_alt}, we show the proton momentum distributions, normalized to $a_1$, for corresponding 1D PIC simulations over a scan in $a_1$. 
Distributions are measured 60 fs 
after the pulse peaks coincide. 
In order to keep the charge separation fields negligible in the low-$a_1$ simulations, we have varied the initial plasma density such that $n_e/n_{c1}=10^{-5} a_1$. 
For $a_1\lesssim35$, the protons favor the $-\hat{\mathbf{x}}$ direction rather than the $+\hat{\mathbf{x}}$ direction; this is because $a_1$ is too low for pulse 1 to stop most of the initially accelerated electrons, so the beat wave motion does not occur. However, for $a_1\gtrsim40$, the distributions agree with the predicted directional bias. As $a_1$ increases further, radiation reaction detraps more particles, causing the momentum distribution to shift leftward. The resulting ion motion could be measured using particle spectrometers.

For the conditions described above, with $a_1=90$, the maximum intensities for both pulses are on the order of $\sim 10^{23}$ W/cm$^2$. Although intensity on this order is achievable in current facilities\cite{Turcu2016HighELI-NP,Radier202210ELI-NP,Yoon2021Realization}, the large spot size means that the required power is out of reach---for a beam waist of 4.25 \textmu m, both pulses require $\sim 80$ PW. From Fig.~\ref{fig:exp_alt}, radiation reaction causes the behavior of the proton momentum distribution function to deviate by $a_1\approx55$ (though without an overall reversal). Achieving $a_1=55$ would require $\sim 30$ PW per laser, in line with proposals for higher-power facilities such as NSF-OPAL \cite{Bromage2019,NSFOPAL}. Increasing the pulse duration would also increase the total energy lost to radiation reaction during the interaction, enhancing the effect's observability at lower power. In addition, while the lasers in an experiment may not be perfectly aligned in space or time, we have observed through a 2D tolerance study that the key dynamics survive misalignments that are small relative to the size and duration of the laser pulses (i.e. offsets on the order of 1 \textmu m or a few fs). Facilities providing tens of PW per laser, which have been proposed \cite{Bromage2019,NSFOPAL,Kawanaka2016} or are currently under construction \cite{Vulcan2020}, could thus conceivably enable a version of this effect to be measured. 

The non-radiative detrapping threshold, Eq.~(\ref{eq:thresh5}), diminishes in accuracy for higher $R_\lambda$ (though it typically stays accurate to within a factor of 2). 
Preliminary investigations with different $a_1$ also suggest that the numerical value of the $r_cw/R_\lambda$ detrapping threshold depends weakly on the pulse amplitude; further analysis is needed to determine the form of that dependence. 
The numerical bounds given should be viewed therefore as first estimates, before configuration-specific simulations are used to narrow down optimal experimental conditions. For the best predictive utility, these configuration-specific simulations should ideally be performed in higher dimensions, to account for any quantitative differences due to the transversely non-uniform laser fields and plasma response. 

In summary, we have shown an effect by which radiation reaction changes the dominant direction of proton motion, mediated by the electron dynamics. The regime in which this occurs is roughly bounded by the following requirements: ${100\gtrsim R_aR_\lambda^2/R_w^{2}}$; $0.75\lesssim r_cw/R_\lambda $; and $v_{ZMF} \gtrsim 1/2$, i.e. $R_\lambda\gtrsim3$. Physically, this means that the temporal dynamics of the two counterpropagating pulses must be insufficient to free most electrons from the interference beat wave, while radiation reaction must be strong enough to do so. 
The phenomenon discovered offers 
a simple way to probe the transition to the radiation-dominated regime.

\begin{acknowledgments}
This work was partially supported by NSF Grant PHY-2308641. The PIC code \textsc{epoch} used in this work was in part funded by the UK EPSRC grants EP/G054950/1, EP/G056803/1, EP/G055165/1 and EP/M022463/1. 
The authors as members of the \textsc{osiris} Consortium acknowledge the use of \textsc{osiris}.
Some of the computing for this project was performed on the Sherlock cluster at Stanford. We would like to thank Stanford University and the Stanford Research Computing Center for providing computational resources and support that contributed to these research results.
\end{acknowledgments}

\section*{Author Declarations}
\subsection*{Conflict of Interest}
The authors have no conflicts to disclose.
\subsection*{Author Contributions}
\textbf{Caleb Redshaw:} Conceptualization (lead); formal analysis (lead); methodology (lead); visualization (lead); writing --- original draft (lead); writing --- review and editing (equal).
\textbf{Matthew R. Edwards:} Conceptualization (supporting); funding acquisition (lead); supervision (lead); visualization (supporting); writing --- review and editing (equal). 

\section*{Data Availability Statement}

The data that support the findings of this study are available from the corresponding author upon reasonable request.

\appendix
\section{Simulation Parameters} \label{app:param}
Table \ref{table:param} lists the parameters used for the simulations corresponding to Figs. \ref{fig:netx}-\ref{fig:exp_alt}. All simulations use a resolution of 40 cells/$\lambda_1$, and in all 2D simulations, the resolution in both directions is the same. 1D simulations use 40 particles per species in each cell in regions where the plasma density is non-zero at the beginning of the simulation, while 2D simulations use 5 particles per species per cell in the same regions. 2D results use a beam waist of 8.5 \textmu m for both pulses at focus.
For all simulations, the initial plasma is uniform, neutral, and composed of electrons and protons, with a thickness of 10 \textmu m and a temperature of 10 eV. The plasma for all simulations has an initial number density of $n_e=10^{-5} a_1 n_{c1}$, where $n_{c1}=m_e\omega_1^2/4\pi e^2$ is the critical density for pulse 1.

\begin{table}[h]
\caption{Physical and Computational Parameters \label{table:param}}
\label{tbl:params}
\begin{ruledtabular}
\begin{tabular}{l c c c c c c c c}
\noalign{\smallskip}
{\bf Fig.} &
Dim.\footnote{All simulations, whether 1D or 2D, calculate all three components of particle velocity.} &
$\lambda_1$ (\textmu m) \footnote{$\lambda_1$ is the wavelength of the pulse incident from the left side of the domain.}&
$a_1$\footnote{$a_1$ is the normalized vector potential of the pulse incident from the left side of the domain. For the 2D simulations, this is defined at focus at the center of the domain.}&
$w_1^L$ \footnote{$w_1^L$ is the laboratory-frame characteristic width of the pulse incident from the left side of the domain, with envelope $g_1(x,t)=e^{-(t-x-\phi_{d,1})^2/(w_1^L)^2}$. $w_1^L$ is normalized by $T_1=\lambda_1/c$.}&

$R_\lambda$\footnote{$R_\lambda = \lambda_2/\lambda_1$.} &
$R_a$\footnote{$R_a = a_2/a_1$.} &
$R_w$\footnote{$R_w=w_2^L/w_1^L$.} 
&
RR?\\ 
\noalign{\smallskip}
\hline
\noalign{\medskip}
\ref{fig:netx}a & 1 & 0.4 & 100 & 6.75  & 3 &  5 & 1 & No\\
\ref{fig:netx}b & 1 & 0.4 & 100 & 6.75  & 3 &  5 & 1 & Yes\\
\ref{fig:netx}c & 1 & 0.4 & 100 & 6.75  & 3 &  [1-10] & 1 & Both\\
\ref{fig:SPM}abc & 1 & 0.4 & 100 & 6.75  & 3 &  5 & 1 & No\\
\ref{fig:beatwave}ab & 1 & 0.4 & 100 & 6.75  & 3 &  5 & 1 & Both\\
\ref{fig:bigscan}a & 1 & 0.4 & 100 & 6.75  & [1-7] & [1-40] & 1 & No\\
\ref{fig:bigscan}b & 1 & 0.4 & 100 & 6.75  & [1-7] & [1-40] & 1 & Yes\\
\ref{fig:bigscan}c & 1 & 0.4 & 100 & 6.75  & [1-7] & [1-40] & 1 & Both\\
\ref{fig:wratio}a & 1 & 0.4 & 100 & 9  & [1-7] & [1-40] & 0.75 & Both\\
\ref{fig:wratio}b & 1 & 0.4 & 100 & 6.75  & [1-7] & [1-40] & 1.33 & Both\\
\ref{fig:2Dexample}a & 2 & 0.4 & 100 & 6.75 & 3 & 5 & 1 & No\\
\ref{fig:2Dexample}b & 2 & 0.4 & 100 & 6.75 & 3 & 5 & 1 & Yes\\
\ref{fig:2Dexample}c & 2 & 0.4 & 100 & 6.75 & 3 & 5 & 1 & Both\\
\ref{fig:exp_alt}a & 1 & 0.267 & [5-130] & 10.1  & 3 & 2.75 & 1 & No\\
\ref{fig:exp_alt}b & 1 & 0.267 & [5-130] & 10.1  & 3 & 2.75 & 1 & Yes\\
 \end{tabular}
\end{ruledtabular}
\end{table}
\bibliography{rr_ions}

\end{document}